\documentclass[twocolumn,journal]{IEEEtran}
\usepackage[T1]{fontenc}
\usepackage{array}
\usepackage{multirow}
\usepackage{graphicx}
\usepackage[unicode=true,
 bookmarks=true,bookmarksnumbered=true,bookmarksopen=true,bookmarksopenlevel=1,
 breaklinks=false,pdfborder={0 0 0},pdfborderstyle={},backref=false,colorlinks=false]
 {hyperref}
\hypersetup{pdftitle={Your Title}, 
 pdfauthor={Your Name},
 pdfpagelayout=OneColumn, pdfnewwindow=true, pdfstartview=XYZ, plainpages=false}

\makeatletter

\providecommand{\tabularnewline}{\\}

 \let\oldforeign@language\foreign@language
 \DeclareRobustCommand{\foreign@language}[1]{%
   \lowercase{\oldforeign@language{#1}}}

\usepackage[caption=false,font=footnotesize]{subfig}

\makeatother

\begin{document}

\title{Wind models and cross-site interpolation for the refugee reception
islands in Greece}

\author{Harris~V.~Georgiou~(MSc, PhD)\thanks{Harris~Georgiou is an associate post-doc researcher with the Data
Science Lab (InfoLab), School of Informatics \& Communications Technologies
(ICT), University of Piraeus (UniPi), Greece.\protect \\
E-mail: \protect\href{mailto:harris@xgeorgio.info}{harris@xgeorgio.info}
\textendash{} URL: \protect\href{http://xgeorgio.info}{http://xgeorgio.info}}}

\IEEEspecialpapernotice{Part I: Time frame Oct.2015\ \textendash \ Jan.2016\\
Last updated: 25-July-2017}

\markboth{Ref.No: HG/DA.0725.01v1 \textendash{} Licensed under Creative Commons
(BY-NC-SA) 4.0~\copyright~2017 Harris Georgiou}{}
\maketitle
\begin{abstract}
In this study, the wind data series from five locations in Aegean
Sea islands, the most active `hotspots' in terms of refugee influx
during the Oct/2015 - Jan/2016 period, are investigated. The analysis
of the three-per-site data series includes standard statistical analysis
and parametric distributions, auto-correlation analysis, cross-correlation
analysis between the sites, as well as various ARMA models for estimating
the feasibility and accuracy of such spatio-temporal linear regressors
for predictive analytics. Strong correlations are detected across
specific sites and appropriately trained ARMA(7,5) models achieve
1-day look-ahead error (RMSE) of less than 1.9 km/h on average wind
speed. The results show that such data-driven statistical approaches
are extremely useful in identifying unexpected and sometimes counter-intuitive
associations between the available spatial data nodes, which is very
important when designing corresponding models for short-term forecasting
of sea condition, especially average wave height and direction, which
is in fact what defines the associated weather risk of crossing these
passages in refugee influx patterns.
\end{abstract}

\begin{IEEEkeywords}
weather analytics, predictive modeling, correlation analysis, ARMA,
refugee influx forecasting
\end{IEEEkeywords}

\IEEEpeerreviewmaketitle{}

\section{Introduction}

\IEEEPARstart{D}{uring} the months between October 2015 and March
2016, Greece has witnessed an unprecedented influx of refugees across
the sea borders with Turkey. Tens or even hundreds of boats, each
carrying 50 people or more, were landing on a daily basis throughout
the Greek islands that are closest to the Turkish coastline. One of
the most important factors in forecasting the influx rate within a
short time frame of 24-48 hours is the weather conditions in these
sea passages, specifically wind intensity and direction, as these
are directly associated with the severity of sea condition and, hence,
the danger involved in the crossing. The most lethal events of boats
capsizing and sinking happened mostly in days and nights when winds
were growing in power or shortly after steady strong winds had already
produced high waves in these areas.

Unfortunately, weather conditions is only one of the factors affecting
the forecasting of influx rates. Compared to longer trips, e.g. between
Libya and Italy, the case of Greece involves a relatively short trip
of 5-7 n.m. and a couple of hours at most, for a fully loaded boat
and working engine. This means that even in bad weather, some groups
of refugees dared to come across (sometimes forced to by the smugglers)
even in very risky conditions. The result was 3,771 registered deaths
and many more missing from hundreds of capsizing and sinking events
during 2015 alone. According to the United Nations High Commissioner
for Refugees (UNHCR) \cite{UNHCR-data3-url}, the International Organization
for Migration (IOM) \cite{IOM-data-url} and the Medecins Sans Frontieres
(MSF) \cite{MSF-rep-url}, during 2015 more than a million people
reached Europe from Turkey and North Africa, seeking safety and asylum.

In order to better organize Search \& Rescue (SAR) sea operations
and logistical support for the humanitarian relief teams at the first
reception islands, it is crucial that some level of influx forecasting
is available. Indeed, previous works have shown that this is a realistic
goal that can be addressed with data-driven methods in the short-term
\cite{influxsmug_2016,influxanal_arxiv_2016,influxanal_safeevros_2016},
regardless of the more general factors and political decisions that
affect the refugee flows in the long-term. Hence, it is extremely
important to have appropriate tools for up-to-date weather forecasting
that uses local sources and weather stations, specifically for wind
speed and direction at these sea passages. The idea is to have these
forecasts available as inputs for the influx prediction modules, together
with proper time-series analysis on both the weather data as well
as the influx data itself.

It should be noted that the alternative of using full weather data
from sources like the Greek Meteorological Service (EMY) or NOAA is
good for postmortem analysis and training datasets, but inappropriate
in actual deployment mode for three main reasons: (a) these data are
usually massive, small mobile devices cannot parse them, (b) they
require reliable Internet access for large downloads, (c) they are
available mostly as forecasts 3-6 hours beforehand, instead of processing
real-time data from local weather stations. Therefore, having simple
prediction models for winds that can run fast and offline is extremely
useful if they are to be deployed in actual SAR operations.

In this study, five areas of first-reception refugee influx `hotspots'
is used as the baseline for simple predictive modeling in the Aegean
Sea. More specifically, local weather stations in the islands of Lesvos
(Petra and Thermi), Chios, Samos and Kos, the five most active sea
passages and landing areas during the specific high-peak period, are
used as input data for analyzing the statistics of winds (average
speed, gust, direction). Furthermore, these five input sources are
used in designing simple linear regressors for covering other intermediate
areas of interest via cross-site interpolation. Experimental results
are presented separately for each location, as well as test cases
of two linear regressors, one for near-spot and one for far-spot wind
prediction.

\section{Material and Data Overview}

\subsection{Weather data series}

The data used in this study are a subset of the 2015-2016 weather
data feeds from Meteo.gr, an open-access Internet portal \cite{Meteo-data-url}
maintained by the National Observatory of Greece for providing weather
data from ground stations. These data were collected and aggregated
in the special dataset GR-RWL1-O15J16 \cite{AegeanW-dataset-url},
already used in other works for training predictive models that combine
localized weather with refugee influx time series.

\begin{figure*}[tbph]
\begin{centering}
\textsf{\includegraphics[width=17cm]{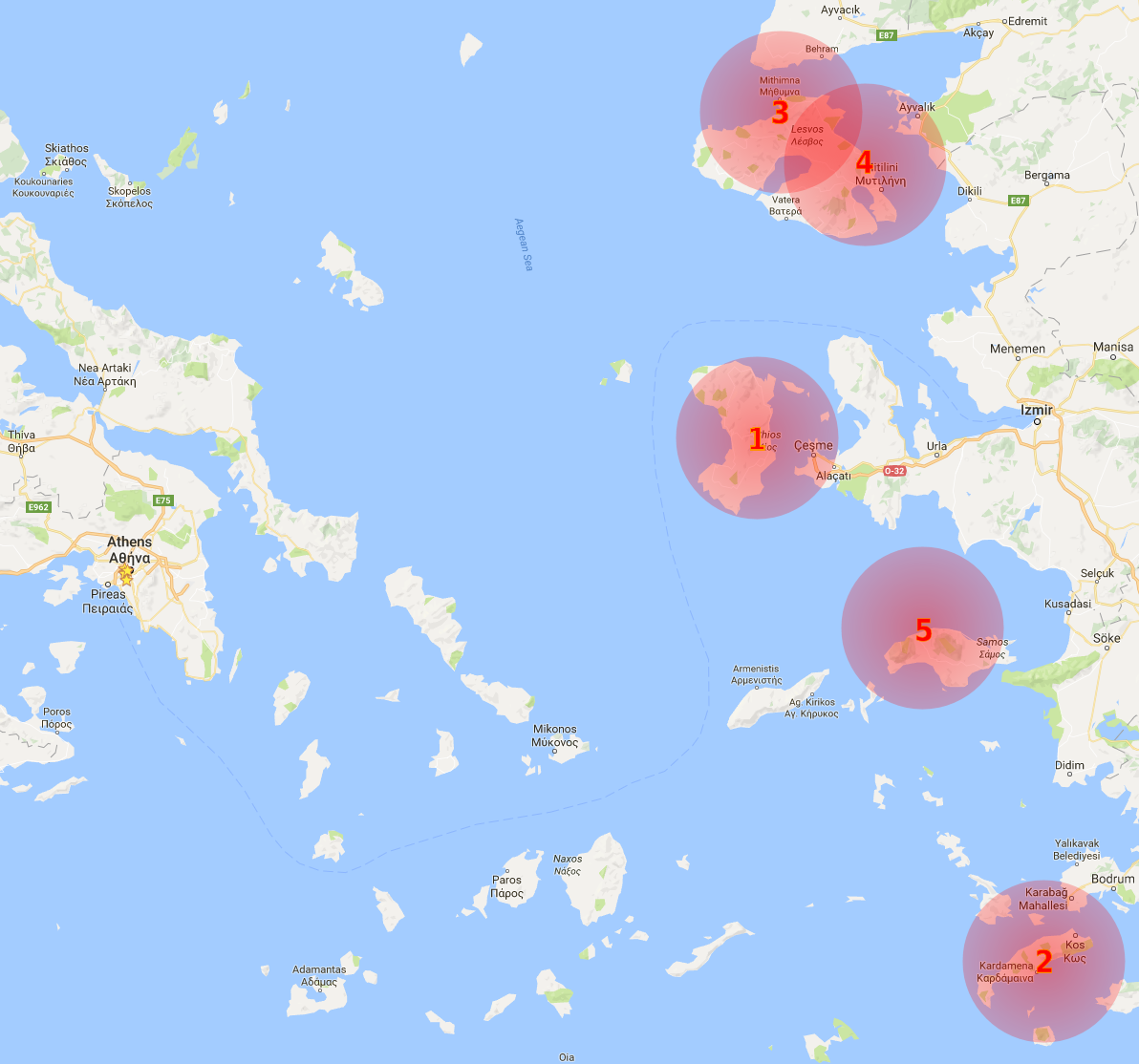}}
\par\end{centering}
\caption{\label{fig:Weather-locations-Greece}Five locations of refugee sea
passages in the Aegean Sea islands, with local weather stations and
detailed wind data (Oct.2015-Jan.2016): (1) Chios, (2) Kos, (3) Lesvos/Petra,
(4) Lesvos/Thermi, (5) Samos.}
\end{figure*}

As mentioned above, five locations with local weather stations were
used as the testbed for this study, specifically in the islands of
Chios, Kos, Lesvos (Petra and Thermi) and Samos. These were the five
most active sea passages and landing areas during the specific high-peak
period (1-Oct-2015 to 31-Jan-2016), associated with more than 85\%
of the total influx in the Aegean Sea, and they are used as input
data for analyzing the statistics of winds, namely the daily values
of average speed, gust and direction. Figure \ref{fig:Weather-locations-Greece}
shows the islands and the approximate locations (center of circle)
of the local weather stations used as data feeds in this study.

\subsection{Software packages and hardware}

The main software packages that were used in this study were:
\begin{itemize}
\item Mathworks MATLAB v8.6 (R2015b), including: Signal Processing Toolbox,
System Identification Toolbox, Statistics \& Machine Learning Toolbox
\cite{Matlab-url}.
\item Additional toolboxes for MATLAB (own \& third-party) for specific
algorithms, as referenced later on in the corresponding sections.
\item WEKA v3.7.13 (x64). Open-Source Machine Learning Suite \cite{Weka-url}.
\item Spreadsheet applications: Microsoft Excel 2007, LibreOffice Sheet
5.1 (x64).
\item Custom-built programming tools in Java and C/C++ for data manipulation
(import/export).
\end{itemize}
The data experiments and processing were conducted using: (a) Intel
i7 quad-core @ 2.0 GHz / 8 GB RAM / MS-Windows 8.1 (x64), and (b)
Intel Atom N270 dual-core @ 1.6 GHz / 2 GB RAM / Ubuntu Linux 16.04
LTS (x32).

\section{Methods and Algorithms}

The statistical and frequency properties of the daily arrivals data
series were analyzed via pairwise correlation and full system identification,
specifically by Auto-Regressive Moving Average (ARMA) approximations,
as described below.

\subsection{Auto-correlation analysis}

Pairwise correlation produces a quantitative metric for the statistical
dependencies between values of two data series at different lags.
In the case when a single data series is compared to itself, the \emph{auto-correlation}
corresponds to the statistical dependencies between subsequent values
of the same series \cite{Hsu-SigSys-1995,Hamming-filters-1989}. Hence,
value pairs with high correlation correspond to regular patterns in
the series, i.e., encode periodicity at smaller or larger scales,
according to the selected lag:

\begin{equation}
R_{yy}\left(t_{1},t_{2}\right)=R\left(k\right)=\frac{\mathrm{E}\left[\left(y\left(t\right)-\mu_{t}\right)\cdot\left(y\left(t+k\right)-\mu_{t+k}\right)\right]}{\sigma_{t}\cdot\sigma_{t+k}}\label{eq:autocorr-def}
\end{equation}

where $y\left(t\right)$ is the time series, $\mu_{t}$ and $\sigma_{t}$
are the mean and standard deviation, $k$ is the lag and $R\left(k\right)$
is the correrponding auto-correlation vector of length $2k+1$. In
this study, the wind variables of average speed and gust were analyzed
via auto-correlation with a lag limit of $k\pm122$ against the current
day, i.e., within the full 123-days time frame available (four months).

\subsection{ARMA system identification}

More generic and powerful than auto-correlation or standard linear
regression alone, the \emph{Auto-Regressive Moving Average} (ARMA)
model is the standard approach for describing any linear digital filter
or signal generator in the time domain. It is essentially a combination
of an auto-correlation component that relates the current outputs
to previous ones and a smoothing component than averages the inputs
over a fixed-size window.

The typical linear ARMA model is described as \cite{Hamming-filters-1989,Porat-signals-1994,Ther-RndSig-1992,Han-ARMA-1980}:

\begin{equation}
A_{m}\left(z\right)\ast\overrightarrow{y}\left(t\right)=B_{k}\left(z\right)\ast\overrightarrow{u}\left(t\right)+e\left(t\right)\label{eq:ARMA-generic}
\end{equation}
where $\overrightarrow{u_{k}}\left(t\right)$ is the input vector
of size $k$ at time step $t$, $\overrightarrow{y}\left(t\right)$
is the output vector of size $m$ (i.e., the current plus the $m-1$
previous ones), $B_{k}\left(z\right)$ is the convolution kernel for
the inputs, $A_{m}\left(z\right)$ is the convolution kernel for the
outputs and $e\left(t\right)$ is the residual model error. Normally,
$A_{m}\left(z\right)$ and $B_{k}\left(z\right)$ are vectors of scalar
coefficients that can be fixed, if the model is static, or variable,
if the model is adaptive (constantly ``retrained''). 

Both coefficient vectors, as well as their sizes, are subject to optimization
of the model design according to some criterion, which typically is
the minimization of the residual error $e\left(t\right)$. In practice,
this is defined as $e\left(t\right)=\left\Vert \hat{y}\left(t\right)-y\left(t\right)\right\Vert _{2}^{2}$,
where $\left\Vert .\right\Vert _{2}$ is the standard Euclidean norm,
$\hat{y}\left(t\right)$ is the ARMA-approximated output and $y\left(t\right)$
is the true (measured) process output. The sizes $m$ and $k$ are
the \emph{orders} of the model and they are usually estimated either
by information-theoretic algorithms \cite{Hamming-filters-1989,Porat-signals-1994,Han-ARMA-1980}
or by exploiting known properties (if any) of the generating process,
e.g. with regard to its periodicity. Such a model is described as
ARMA($m$,$k$), where AR($m$) is the auto-regressive component and
MA($k$) is the moving-average component. 

In approximation form, expanding the convolutions and estimating the
current output $\hat{y}\left(t\right)$, the analytical form of Eq.\ref{eq:ARMA-generic}
is:

\begin{equation}
\hat{y}\left(t\right)=\sum_{i=1}^{m}\left(a_{i}\cdot y\left(t-i\right)\right)+\sum_{j=0}^{k}\left(b_{j}\cdot x\left(t-j\right)\right)+e\left(t\right)\label{eq:ARMA-analytical}
\end{equation}

The error term $e\left(t\right)$ in Eq.\ref{eq:ARMA-analytical}
can also be expanded to multiple terms of a separate convolution kernel,
similarly to $A_{m}\left(z\right)$ and $B_{k}\left(z\right)$, but
it is most commonly grouped into one scalar factor, i.e., with an
order of one. In such cases, the model and be described as ARMA($m$,$k$,$q$)
where $q>1$ is the order of the convolution kernel for the error
term.

When applied to a signal generated by a process of unknown statistical
properties, an ARMA approximation of it reveals a variety of important
properties regarding this process. In practice, the (estimated) order
$m$ of the AR component shows how strong the statistical coupling
is between subsequent outputs, while the order $k$ of the MA component
shows the ``memory'' of the process, i.e., how far in the past inputs
the process ``sees'' in order to produce the current output.

\subsection{Interpolation for missing values}

In the selected time frame, the weather dataset for Samos contained
one day of missing data (10-Jan-2016) due to maintenance of the weather
station there. In order to fill-in the missing data, interpolation
was applied separately for each target wind variable, i.e., average
speed, gust and direction. 

More specifically, moving average of two and four points (missing
point in the middle) were tested, as well as cubic spline interpolation
(QS) \cite{Schaum-MathTabl-2012}, as shown in Table \ref{tab:Samos-missing-value}
for the `average wind speed' parameter. The results from leave-one-out
cross-validation tests across the entire Samos data series per-dimension
confirmed, as expected, that QS is the most resilient to interpolation
errors (RMSE), compared to the MA(2) and MA(4) methods. Moreover,
using more than two points immediately adjacent to the missing one
produces slightly larger interpolation error when MA is employed. 

\begin{table}[htbp]
\caption{\label{tab:Samos-missing-value}Interpolation values for missing data
(10-Jan-2016) and leave-one-out cross-validation error over the entire
Samos `average wind speed' data series (123 days).}
\centering{}%
\begin{tabular}{|c|c|c|c|}
\hline 
Wind Speed (km/h) & Method & Value & RMSE\tabularnewline
\hline 
\hline 
\multirow{3}{*}{average} & MA(2) & 3.00 & 3.25\tabularnewline
\cline{2-4} 
 & MA(4) & 3.85 & 3.52\tabularnewline
\cline{2-4} 
 & \textbf{QS} & 2.57 & \textbf{2.97}\tabularnewline
\hline 
\multirow{3}{*}{gust} & MA(2) & 25.75 & 9.81\tabularnewline
\cline{2-4} 
 & MA(4) & 35.80 & 10.67\tabularnewline
\cline{2-4} 
 & \textbf{QS} & 21.73 & \textbf{9.33}\tabularnewline
\hline 
\end{tabular}
\end{table}

These results hint that there may be large short-term fluctuations
present in the data series and higher-order polynomial, QS or other,
has to be employed for accurate interpolation of the missing values.
This is also evident by the fact that the actual interpolated values
from MA(2) in Table \ref{tab:Samos-missing-value} are much closer
to the QS values, measured as overall the most accurate w.r.t. RMSE
(see Eq.\ref{eq:rmse-def}), than the ones produced by MA(4). Hence,
using the simple rule-of-thumb that assumes only small changes from
the previous 24-hour time frame seems to be unsafe in practice for
accurate forecasting.

The QS interpolation used in this case throughout he study is expected
to be adequately accurate, namely $\pm3$ km/h for average wind speed
and about $\pm9.3$ km/h for wind gust, which for the winds scale
for Samos (see next section) translates to less than $\pm1$B for
both `average' and `gust' parameters. Therefore, this single-day interpolated
instance in the Samos data series can be considered as non-intrusive
for the inherent statistics and the validity of the main experimental
protocol.

\section{Experiments and Results}

The following subsection present the analysis and modeling of the
wind data series for the five sites, including (a) statistics per
location, (b) correlation analysis and (c) ARMA for predictive modeling.

\subsection{Statistics per location}

\begin{figure}[htbp]
\begin{centering}
\textsf{\includegraphics[width=8.5cm]{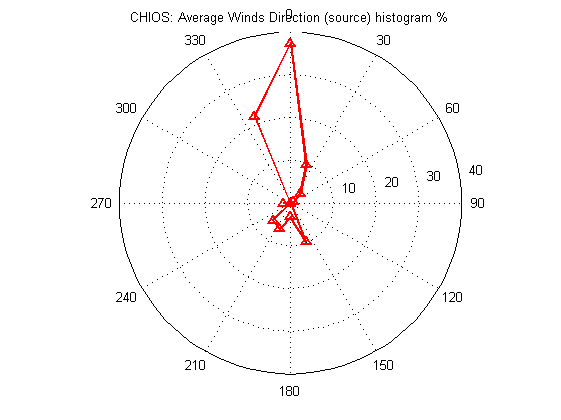}}
\par\end{centering}
\caption{\label{fig:Chios-wdir}Normalized polar histogram of dominant winds
(daily average) for Chios.}
\end{figure}

\begin{figure}[htbp]
\begin{centering}
\textsf{\includegraphics[width=8.5cm]{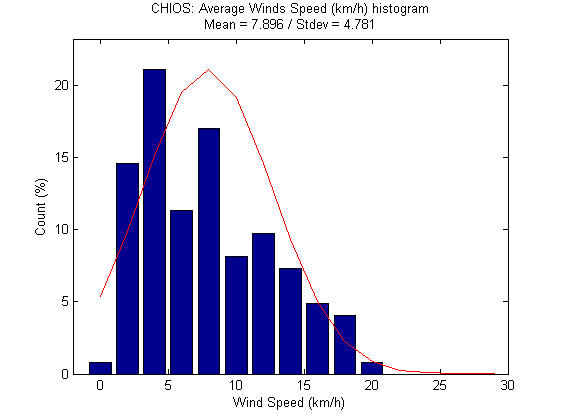}}
\par\end{centering}
\caption{\label{fig:Chios-wavg}Normalized histogram of average wind speed
for Chios.}
\end{figure}

\begin{figure}[htbp]
\begin{centering}
\textsf{\includegraphics[width=8.5cm]{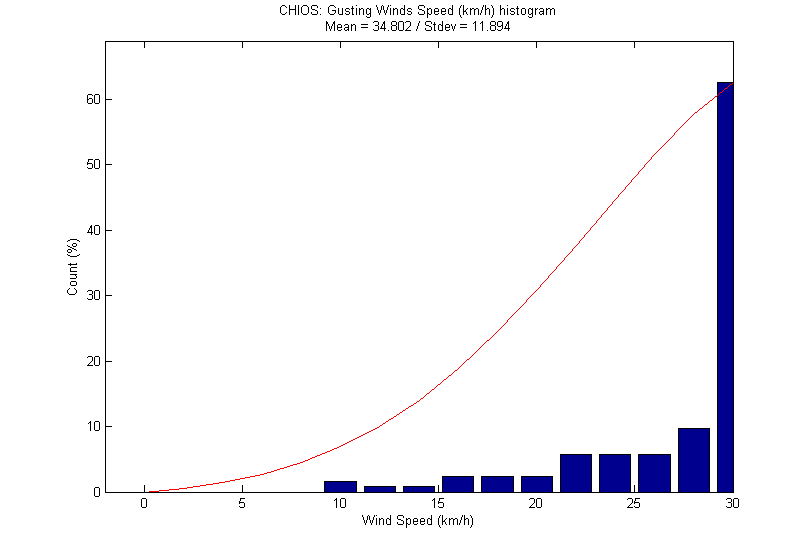}}
\par\end{centering}
\caption{\label{fig:Chios-wgust}Normalized histogram of wind gusts for Chios.}
\end{figure}

\begin{figure}[htbp]
\begin{centering}
\textsf{\includegraphics[width=8.5cm]{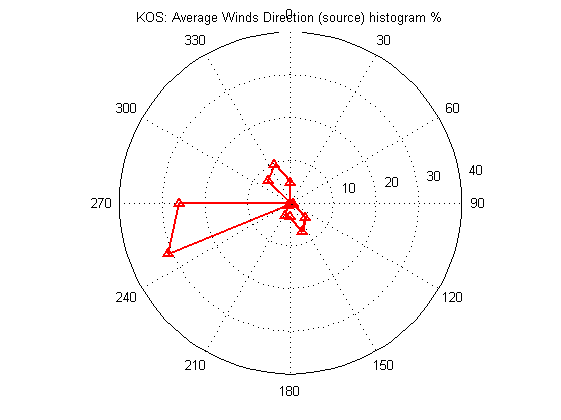}}
\par\end{centering}
\caption{\label{fig:Kos-wdir}Normalized polar histogram of dominant winds
(daily average) for Kos.}
\end{figure}

\begin{figure}[htbp]
\begin{centering}
\textsf{\includegraphics[width=8.5cm]{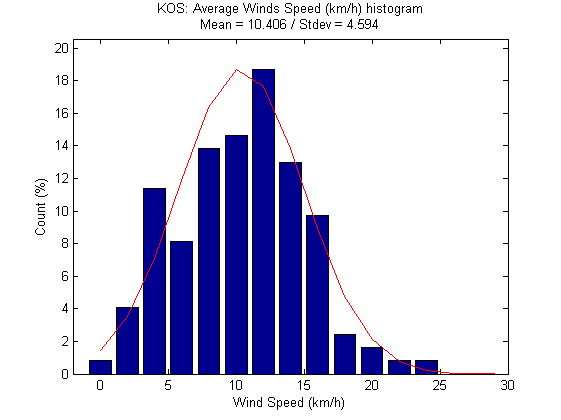}}
\par\end{centering}
\caption{\label{fig:Kos-wavg}Normalized histogram of average wind speed for
Kos.}
\end{figure}

\begin{figure}[htbp]
\begin{centering}
\textsf{\includegraphics[width=8.5cm]{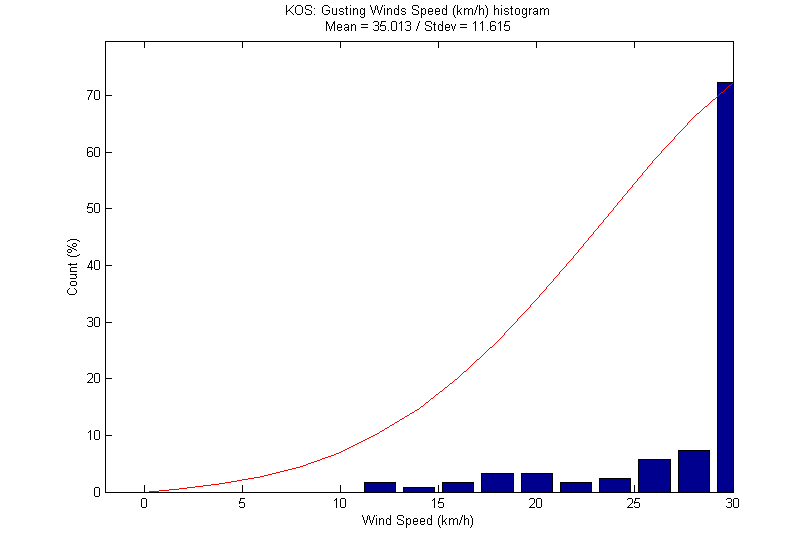}}
\par\end{centering}
\caption{\label{fig:Kos-wgust}Normalized histogram of wind gusts for Kos.}
\end{figure}

\begin{figure}[htbp]
\begin{centering}
\textsf{\includegraphics[width=8.5cm]{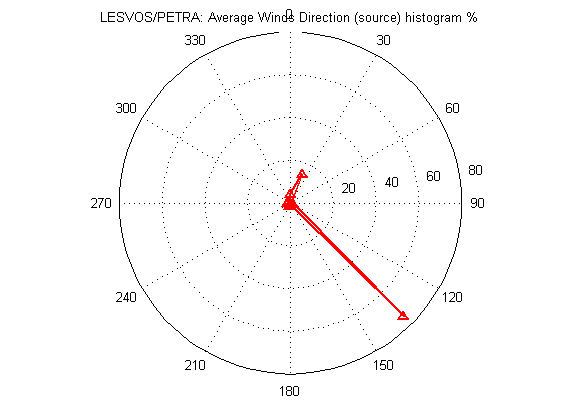}}
\par\end{centering}
\caption{\label{fig:LesvosP-wdir}Normalized polar histogram of dominant winds
(daily average) for Lesvos/Petra.}
\end{figure}

\begin{figure}[htbp]
\begin{centering}
\textsf{\includegraphics[width=8.5cm]{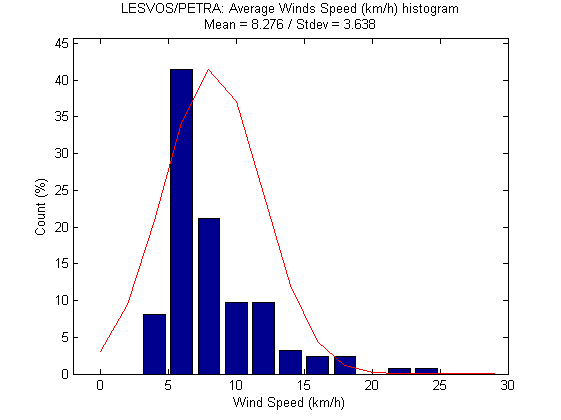}}
\par\end{centering}
\caption{\label{fig:LesvosP-wavg}Normalized histogram of average wind speed
for Lesvos/Petra.}
\end{figure}

\begin{figure}[htbp]
\begin{centering}
\textsf{\includegraphics[width=8.5cm]{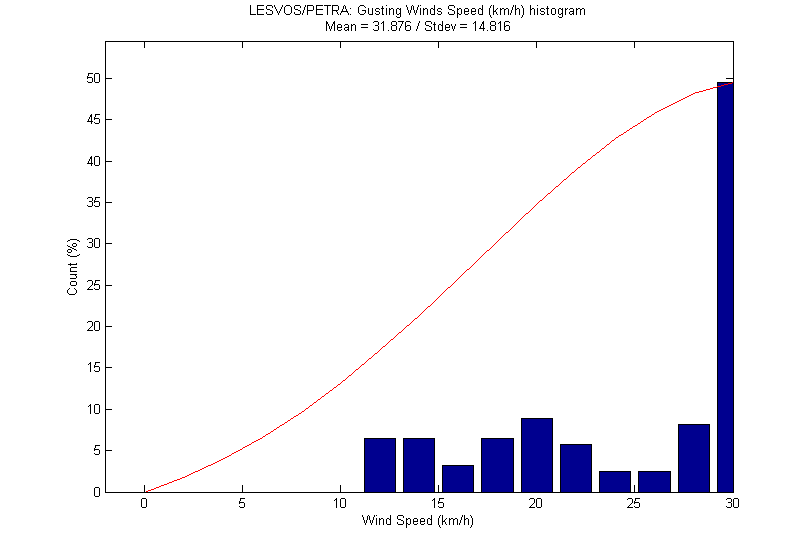}}
\par\end{centering}
\caption{\label{fig:LesvosP-wgust}Normalized histogram of wind gusts for Lesvos/Petra.}
\end{figure}

\begin{figure}[htbp]
\begin{centering}
\textsf{\includegraphics[width=8.5cm]{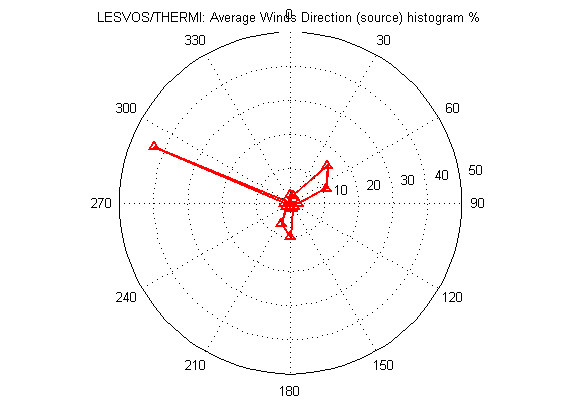}}
\par\end{centering}
\caption{\label{fig:LesvosT-wdir}Normalized polar histogram of dominant winds
(daily average) for Lesvos/Thermi.}
\end{figure}

\begin{figure}[htbp]
\begin{centering}
\textsf{\includegraphics[width=8.5cm]{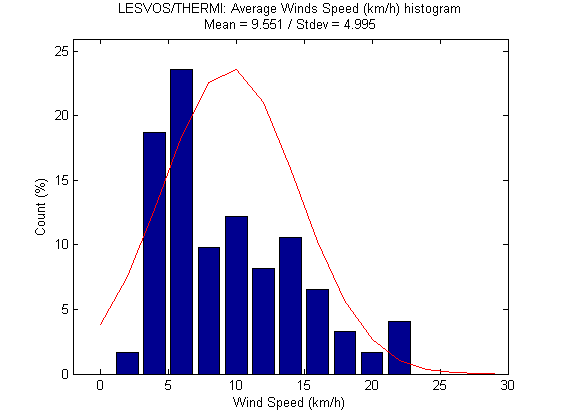}}
\par\end{centering}
\caption{\label{fig:LesvosT-wavg}Normalized histogram of average wind speed
for Lesvos/Thermi.}
\end{figure}

\begin{figure}[htbp]
\begin{centering}
\textsf{\includegraphics[width=8.5cm]{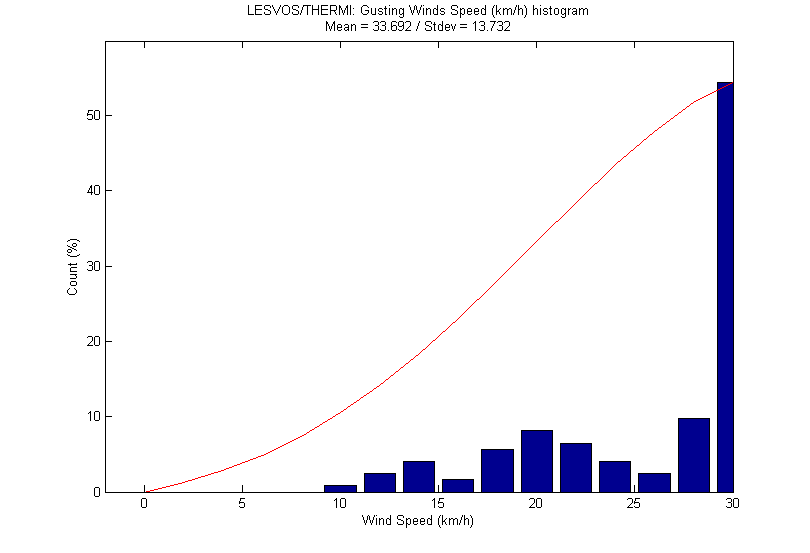}}
\par\end{centering}
\caption{\label{fig:LesvosT-wgust}Normalized histogram of wind gusts for Lesvos/Thermi.}
\end{figure}

\begin{figure}[htbp]
\begin{centering}
\textsf{\includegraphics[width=8.5cm]{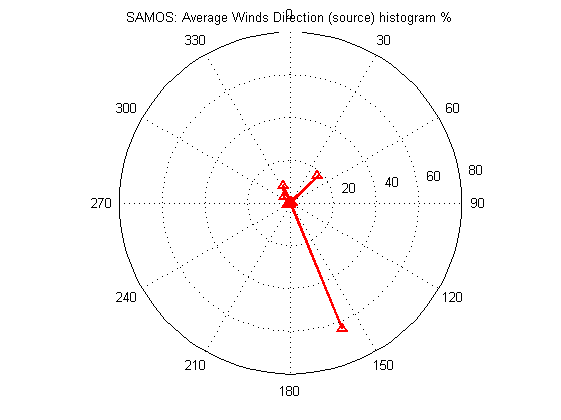}}
\par\end{centering}
\caption{\label{fig:Samos-wdir}Normalized polar histogram of dominant winds
(daily average) for Samos.}
\end{figure}

\begin{figure}[htbp]
\begin{centering}
\textsf{\includegraphics[width=8.5cm]{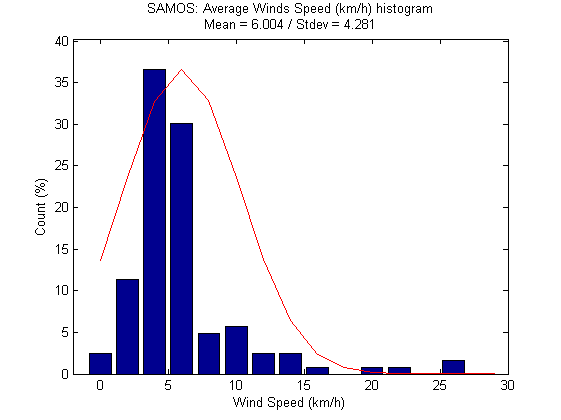}}
\par\end{centering}
\caption{\label{fig:Samos-wavg}Normalized histogram of average wind speed
for Samos.}
\end{figure}

\begin{figure}[htbp]
\begin{centering}
\textsf{\includegraphics[width=8.5cm]{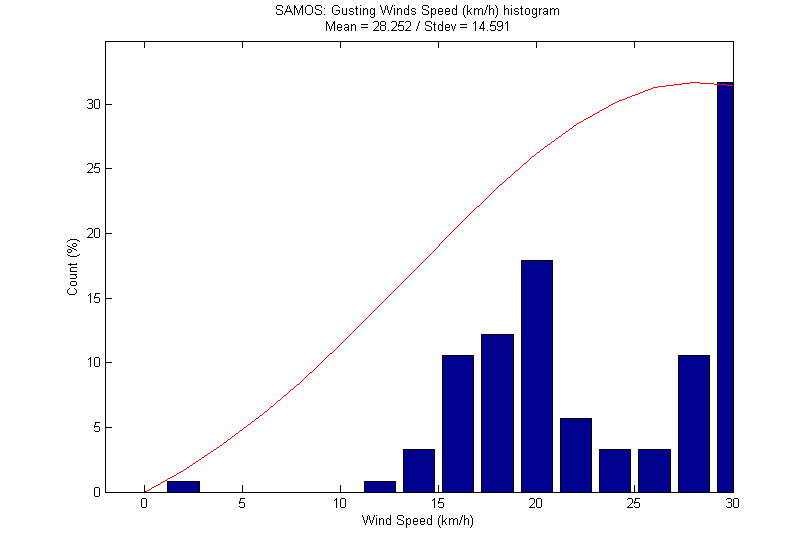}}
\par\end{centering}
\caption{\label{fig:Samos-wgust}Normalized histogram of wind gusts for Samos.}
\end{figure}

The following figures present the statistics of wind direction, average
speed and gusts for the five sites. Figures \ref{fig:Chios-wdir},
\ref{fig:Kos-wdir}, \ref{fig:LesvosP-wdir}, \ref{fig:LesvosT-wdir}
and \ref{fig:Samos-wdir} illustrate the normalized polar histograms
of dominant winds (daily average). Figures \ref{fig:Chios-wavg},
\ref{fig:Kos-wavg}, \ref{fig:LesvosP-wavg}, \ref{fig:LesvosT-wavg}
and \ref{fig:Samos-wavg} illustrate the normalized histograms of
the average speed. Figures \ref{fig:Chios-wgust}, \ref{fig:Kos-wgust},
\ref{fig:LesvosP-wgust}, \ref{fig:LesvosT-wgust} and \ref{fig:Samos-wgust}
illustrate the normalized histogram of gusts. All statistics are for
the time frame used in this study, i.e., 1-Oct-2015 to 31-Jan-2016
(123 days).

\subsection{Correlation analysis }

The following figures present the pairwise auto-correlation analysis
of wind average speed and gust for the five sites. In each plot, the
entire dates range is used (1-Oct-2015 to 31-Jan-2016) in one-step
lags. The `urand' diagonal (dotted) is displayed as guideline for
comparison against a Gaussian random walk with mean and standard deviation
same as the corresponding gust data series.

\begin{figure}[htbp]
\begin{centering}
\textsf{\includegraphics[width=8.5cm]{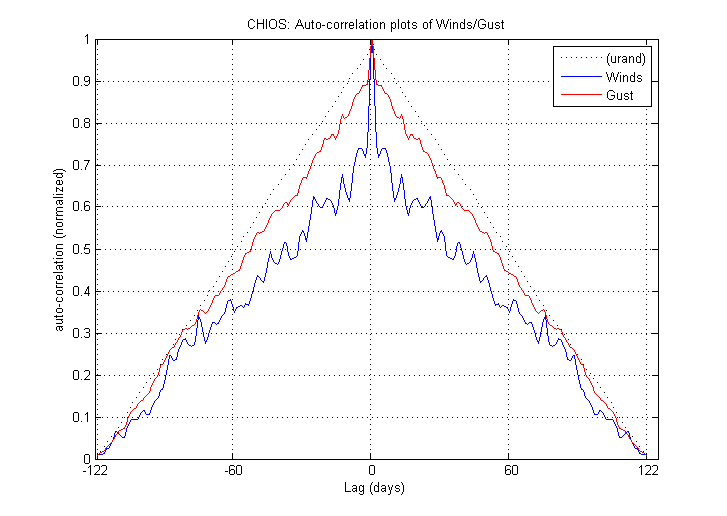}}
\par\end{centering}
\caption{\label{fig:Chios-winds-autocorr}Auto-correlation plot (normalized)
of wind average and gust for Chios. }
\end{figure}

\begin{figure}[htbp]
\begin{centering}
\textsf{\includegraphics[width=8.5cm]{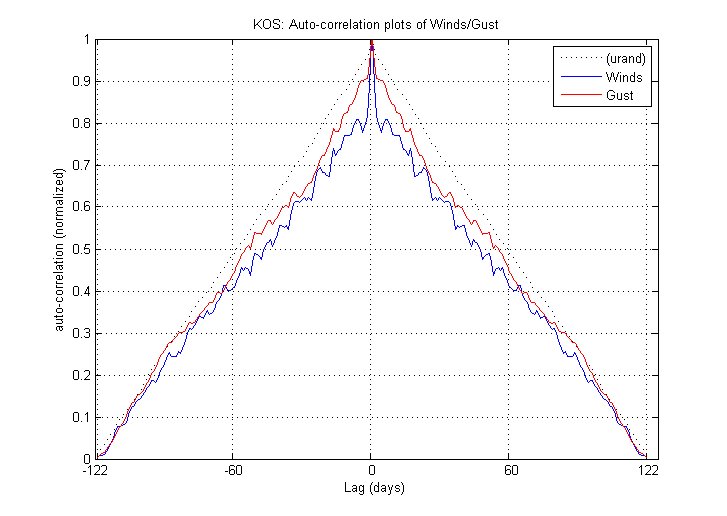}}
\par\end{centering}
\caption{\label{fig:Kos-winds-autocorr}Auto-correlation plot (normalized)
of wind average and gust for Kos.}
\end{figure}

\begin{figure}[htbp]
\begin{centering}
\textsf{\includegraphics[width=8.5cm]{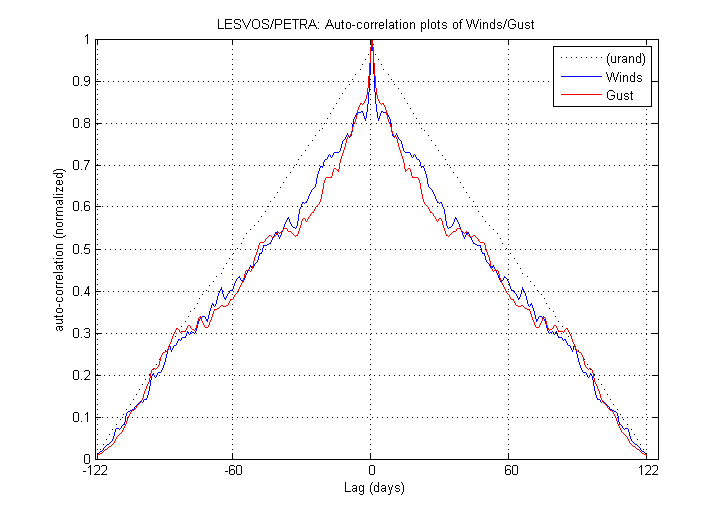}}
\par\end{centering}
\caption{\label{fig:LesvosP-winds-autocorr}Auto-correlation plot (normalized)
of wind average and gust for Lesvos/Petra.}
\end{figure}

\begin{figure}[htbp]
\begin{centering}
\textsf{\includegraphics[width=8.5cm]{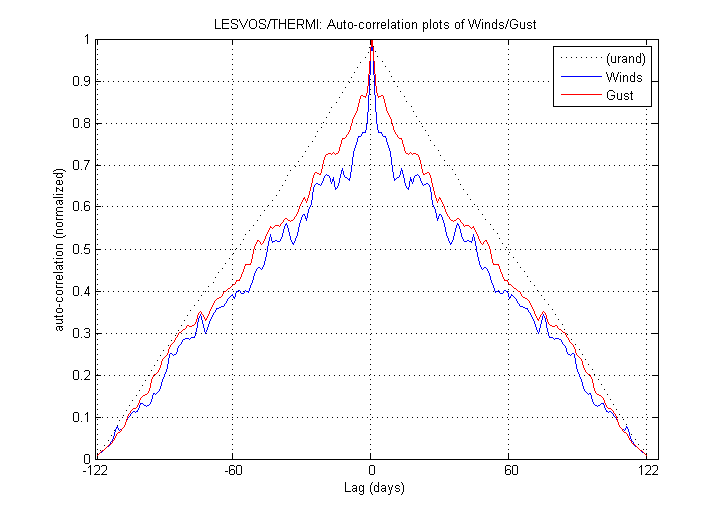}}
\par\end{centering}
\caption{\label{fig:LesvosT-winds-autocorr}Auto-correlation plot (normalized)
of wind average and gust for Lesvos/Thermi.}
\end{figure}

\begin{figure}[htbp]
\begin{centering}
\textsf{\includegraphics[width=8.5cm]{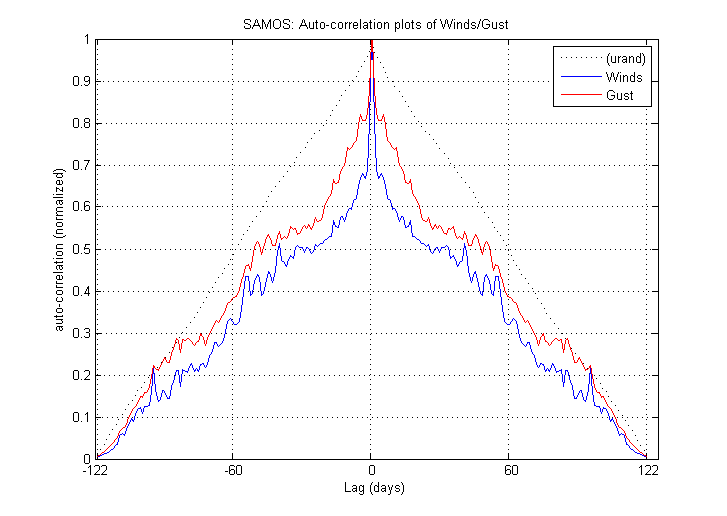}}
\par\end{centering}
\caption{\label{fig:Samos-winds-autocorr}Auto-correlation plot (normalized)
of wind average and gust for Samos.}
\end{figure}

Table \ref{tab:Winds-corr} presents the pairwise correlations between
the three wind data series per site. Statistical significance confirmation
is indicated with bold numbers for $a\leq0.05$ and plain numbers
for $a\leq0.1$, otherwise italics indicate non-significant result
for correlation. The same notation is used in Table \ref{tab:Winds-loc-corr},
presenting the cross-site correlations for wind average speed.

\begin{table}[htbp]
\caption{\label{tab:Winds-corr}Correlations between wind variables for all
five sites (see text for details).}
\centering{}%
\begin{tabular}{|c|c|c|c|}
\hline 
Site & avg/gust & avg/dir & gust/dir\tabularnewline
\hline 
\hline 
\multirow{1}{*}{Chios} & \textbf{0.775} & 0.519 & \emph{0.086}\tabularnewline
\hline 
\multirow{1}{*}{Kos} & \textbf{0.642} & \emph{0.410} & \textbf{0.685}\tabularnewline
\hline 
\multirow{1}{*}{Lesvos/P} & \textbf{0.856} & \emph{0.022} & \emph{0.319}\tabularnewline
\hline 
\multirow{1}{*}{Lesvos/T} & \textbf{0.739} & \textbf{0.837} & \emph{0.452}\tabularnewline
\hline 
\multirow{1}{*}{Samos} & \textbf{0.830} & \emph{0.027} & \emph{0.176}\tabularnewline
\hline 
\end{tabular}
\end{table}

\begin{table}[htbp]
\caption{\label{tab:Winds-loc-corr}Cross-site correlations for wind average
speed (see text for details).}
\centering{}%
\begin{tabular}{|c|c|c|c|c|}
\hline 
Chios & Kos & Lesvos/P & Lesvos/T & Samos\tabularnewline
\hline 
\hline 
1 & \textbf{0.250} & \textbf{0.636} & \textbf{0.884} & 0.168\tabularnewline
\hline 
 & 1 & \textbf{0.277} & \textbf{0.225} & \textbf{0.216}\tabularnewline
\hline 
 &  & 1 & \textbf{0.672} & 0.166\tabularnewline
\hline 
 &  &  & 1 & \textbf{0.204}\tabularnewline
\hline 
 &  &  &  & 1\tabularnewline
\hline 
\end{tabular}
\end{table}

\subsection{ARMA for predictive modeling}

As describe above, ARMA can provide model identification for analysis
and/or prediction. In this study, various AR, MA and ARMA designs
were employed for the three wind variables (speed average, gust, direction)
with a focus on forecasting future values from a historic time frame.

More specifically, in the case of AR a sequence of previous data points
from the \emph{same} series were used to predict future values; in
the case of MA a sequence of previous \emph{and current} data points
from the corresponding series of \emph{other} sites were used to predict
future values; and in the case of ARMA these two approaches are combined.
In practice, the MA part functions as input aggregator, i.e., regression
against the other sites, and the AR part functions as output filter,
i.e., regression against the same site's history.

\begin{figure*}[tbph]
\begin{centering}
\textsf{\includegraphics[width=17cm]{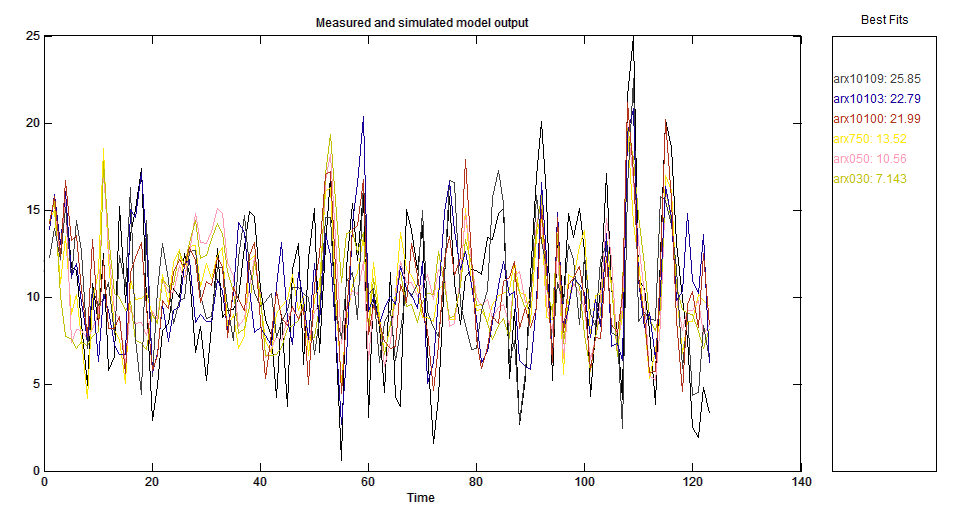}}
\par\end{centering}
\caption{\label{fig:ARX-Kos-4sites-wavg}ARMA(\emph{m},\emph{k}) predictive
modeling for Kos' wind average speed data series (see text for details).}
\end{figure*}

\begin{figure*}[tbph]
\begin{centering}
\textsf{\includegraphics[width=17cm]{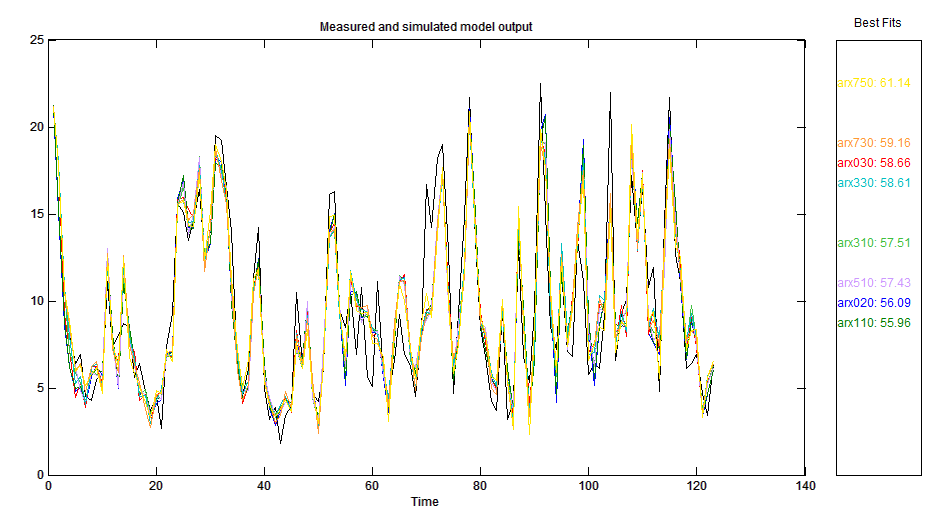}}
\par\end{centering}
\caption{\label{fig:ARX-LesvosT-4sites-wavg}ARMA(\emph{m},\emph{k}) predictive
modeling for Lesvos/Thermi's wind average speed data series (see text
for details).}
\end{figure*}

Figure \ref{fig:ARX-Kos-4sites-wavg} illustrates the results of ARMA
predictive modeling for Kos' wind average speed data series, for various
AR and MA configurations. Even when using an AR component of order
7 or 10 and a MA component of 5 or 10 (historic values from the four
other sites), no model seems to converge accurately in term of prediction.
In contrast, Figure \ref{fig:ARX-LesvosT-4sites-wavg} illustrates
how a similar ARMA (7,5) model achieves very efficient approximation
for the Lesvos/Thermi site. 

The fitness $F$ is a measure of accuracy of the approximation model
and for a specific data series is defined analytically as:

\begin{equation}
F=100\cdot\left(1-\frac{\left\Vert y-\hat{y}\right\Vert _{2}}{\left\Vert y-\bar{y}\right\Vert _{2}}\right)\label{eq:matlab-fitness}
\end{equation}
where $y$ is the true data point, $\hat{y}$ is the model's approximation,
$\bar{y}$ is the data series mean and $\left\Vert .\right\Vert _{2}$
is the standard Euclidean norm. For comparison, RMSE is defined as:

\begin{equation}
err\left(y,\hat{y}\right)_{RMSE}=\sqrt{\frac{1}{n}\sum_{i=1}^{n}\left(y-\hat{y}\right)^{2}}\label{eq:rmse-def}
\end{equation}

From Figure \ref{fig:ARX-LesvosT-4sites-wavg} it is clearly evident
that the ARMA(7,5) approximation is very efficient and successfully
tracks the wind average speed data series very closely, using 7 previous
values from the same site and 4 previous plus the current one from
the other four sites. In other words, the model implements a spatio-temporal
linear regressor with a total of 27 free parameters (trained) to make
a 1-day look-ahead prediction. The exact performance of the model
is $F=62.29(\%)$, FPE=5.064 (Akaike' final prediction error \cite{Akaike_FPE_1999})
and RMSE=1.884 (km/h). The trained ARMA(7,5) model is described below:

\begin{equation}
\begin{array}{cc}
A_{7}\left(z\right) & =1-0.125\cdot z^{-1}-0.081\cdot z^{-2}-0.199\cdot z^{-3}\\
 & +0.231\cdot z^{-4}+0.036\cdot z^{-5}\\
 & -0.059\cdot z^{-6}+0.019\cdot z^{-7}
\end{array}\label{eq:ARMA750-Avec}
\end{equation}
and:

\begin{equation}
\begin{array}{cc}
B_{5,1}\left(z\right) & =0.780-0.105\cdot z^{-1}-0.111\cdot z^{-2}\\
 & -0.039\cdot z^{-3}+0.217\cdot z^{-4}
\end{array}\label{eq:ARMA750-Bvec1}
\end{equation}

\begin{equation}
\begin{array}{cc}
B_{5,2}\left(z\right) & =0.333-0.186\cdot z^{-1}+0.179\cdot z^{-2}\\
 & -0.121\cdot z^{-3}-0.034\cdot z^{-4}
\end{array}\label{eq:ARMA750-Bvec2}
\end{equation}

\begin{equation}
\begin{array}{cc}
B_{5,3}\left(z\right) & =0.083+0.029\cdot z^{-1}+0.033\cdot z^{-2}\\
 & -0.028\cdot z^{-3}-0.008\cdot z^{-4}
\end{array}\label{eq:ARMA750-Bvec3}
\end{equation}

\begin{equation}
\begin{array}{cc}
B_{5,5}\left(z\right) & =0.020-0.069\cdot z^{-1}+0.002\cdot z^{-2}\\
 & +0.058\cdot z^{-3}-0.029\cdot z^{-4}
\end{array}\label{eq:ARMA750-Bvec5}
\end{equation}

In these polynomials, $z^{-n}$ is the delay factor of the kernel,
as described by the analytical form of Eq.\ref{eq:ARMA-analytical}.
Hence, the coefficient of $z^{-n}$ in $A_{7}\left(z\right)$ is essentially
$a_{n}$, i.e., the magnitude for the auto-regressive factor (output)
$n$ days back. The second index $k$ in $B_{5,k}\left(z\right)$
refers to the corresponding site, in the same order as presented above.
Since Lesvos/Thermi ($k=4$) is the target site, all the other sites
are associated to their corresponding polynomials, i.e., Chios with
Eq.\ref{eq:ARMA750-Bvec1}, Kos with Eq.\ref{eq:ARMA750-Bvec2}, Lesvos/Petra
with Eq.\ref{eq:ARMA750-Bvec3} and Samos with Eq.\ref{eq:ARMA750-Bvec5}.

\section{Discussion}

The methods presented in this study fall under the general concept
of regression, i.e., using available values from single or combined
spatio-temporal data series to predict their future evolution. In
this sense, even the missing data interpolation described earlier
for Samos could be used as a very simple and fast approach to do this
in practice. However, it is clear that this is not the optimal approach
in terms of accuracy, as the results from ARMA illustrate later on.
Thus, it is imperative to conduct in-depth statistical and correlation
analysis in each data series separately and in combination, in order
to estimate their spatio-temporal dependencies and inherent complexity,
which are to be used as guidelines for the design of efficient ARMA
or other similar model approximations.

The polar histograms of wind speeds reveal that there are clearly
dominant wind directions in all five sites: for Chios it is North,
for Kos it is West, for Lesvos/Petra it is South-East, for Lesvos/Thermi
it is West/North-West and for Samos it is South/South-East. In the
three later cases the evidence is very strong, showing a very compact
peak towards the source of the dominant wind direction. This is extremely
useful when the wind models are to be associated statistically with
corresponding sea condition models, for example the average wave height,
since the local geographical morphology becomes more or less irrelevant
in these approximations.

In all the sites, the histograms of the wind average speeds present
an almost-Gaussian probability distribution function (pdf) and, thus,
the corresponding mean, standard deviation, etc, for each case can
be considered statistically valid. On the other hand, wind gusts in
all sites is highly skewed if the same wind speed margins are used.
These pdfs may also be approximated by Gaussians, but further investigation
is required in terms of descriptive statistics and parametric distributions,
as it is clearly evident that they include multiple peaks, large `flat'
ranges, etc.

The auto-correlation plots reveal the inherent dependencies between
subsequent values in the data series, separated by specific time lags.
In general, it is expected that non-random signals present a triangular-shaped
plot centered at the zero-lag peak, which exhibits the maximum auto-correlation
value. The larger the deviation from this triangular shape, the larger
the stochastic factors in the signal. In other words, for a perfectly
random signal of pure white noise, the corresponding plot should be
`flat' instead of triangular-shaped. In the plots illustrated here,
the wind average speeds in the five sites present smaller or larger
stochastic properties. In particular, Samos seems to be the most unpredictable
case, with large drops immediately before/after the zero-lag peak,
as well as multiple small peaks in other lag values. on the other
hand, Kos seems to be the most deterministic case, as the plot is
almost perfectly triangular-shaped. It is expected that data series
with large deviations from the `ideal' triangular shape, if approximated
by linear models like ARMA, will require larger convolution kernels,
i.e., higher-order AR components, in order to capture longer historic
sequences.

The correlations between the wind variables, illustrated in Table
\ref{tab:Winds-corr}, reveal that there is verified statistical dependency
($a\leq0.05$) between average speed and gusts, as expected, in all
five sites. Furthermore, there is strong dependency ($a\leq0.05$)
between dominant wind direction and (a) average speed in Lesvos/Thermi
and (b) gust in Kos, as well as (c) average speed in Chios at a lower
level ($a\leq0.10$). In these cases, it is expected that corresponding
sea condition models, especially average wave height and direction,
will be more accurate and useful than in the other sites. 

Furthermore, the cross-correlation results, illustrated in Table \ref{tab:Winds-loc-corr},
reveal that there is a strong association ($a\leq0.05$) between the
sites that are geographically adjacent or adequately near to each
other. In particular, the average wind speed at Chios is strongly
correlated with the two Lesvos sites, especially the one at Thermi.
Looking at the map in Figure \ref{fig:Weather-locations-Greece},
it is clear why: the two islands are near to each other with open
sea between them, especially between Lesvos/Thermi and Chios, which
exhibits the largest cross-correlation value in the Table. There is
also a very strong correlation between the two sites at Lesvos, as
expected. Nevertheless, the link between the two Lesvos sites is still
weaker than the one between Lesvos/Thermi and Chios, probably because
in the first case there is a large geographical obstacle between them
- the large mount of the island of Lesvos itself. It is worth noticing
that the distance across the two sites at Lesvos is about 34 km, while
the distance across the Lesvos/Thermi and Chios sites is about 92
km, more than x2.7 times larger, but entirely over open sea or flat
land. These quantified observations are extremely important when designing
cross-site interpolation models, as only this kind of statistical
analysis can reveal such unexpected and counter-intuitive results
regarding the weight that needs to be assigned to each spatial data
node.

The results from the ARMA model examples more or less confirm the
conclusions drawn from the cross-correlations between sites. The average
wind speed at Kos is difficult to predict from the values of the other
four sites, as Figure \ref{fig:ARX-Kos-4sites-wavg} illustrates,
because their cross-correlations are marginally around 0.25 even for
the closest one (Samos). This example was selected specifically to
show that the most remote sites in terms of spatial distribution is
the most difficult to forecast via linear predictive modeling like
ARMA. There is still some correlation present between the sites over
large geographical areas, primarily because the Aegean Sea is enclosed
from three sides and spans over a basin of roughly 280 x 500 km (main
area). This makes it very rare to have drastically different wind
conditions over its islands, except in areas with specific geographical
features, as in the case of the narrow `closed' sea passage north
of the Lesvos/Petra site, compared to the more `open' passage east
of the Lesvos/Thermi site.

The ARMA(7,5) model detailed for the Lesvos/Thermi site is a typical
example of how spatio-temporal linear regressors can achieve very
accurate predictions in the short term. In particular, the average
wind speed that is modeled can be approximated with an expected error
(RMSE) of less than $\pm1.9$ km/h, which is almost 37\% better than
the error achieved by the cubic spline interpolation (QS) as described
for the Samos data series, despite the fact that in the second case
the interpolation is for an intermediate point (i.e., both sides bounded
by true values) instead of 1-day look-ahead extrapolation (i.e., only
one side bounded). Given the fact that such ARMA models are very simple
to implement, based on vector operations over 30 or less parameters
(trained), it is clearly evident that they can be extremely useful
when analytical weather forecasts like NOAA or computationally intensive
simulation-based models cannot be used in practice, e.g. as part of
a mobile or web application.

\section{Conclusion}

In this study, the three main wind variables (average speed, gust,
direction) were investigated for five main locations that have been
the most active `hotspots' in terms of refugee influx in the Aegean
Sea islands during the Oct/2015 - Jan/2016 period.

The analysis of the three-per-site data series included standard statistical
analysis and parametric distributions (Gaussians), auto-correlation
analysis, as well as cross-correlation analysis between the sites.
Various ARMA models were designed and trained in order to estimate
the feasibility and accuracy of such spatio-temporal linear regressors
for predictive analytics, compared also with standard moving average
and cubic spline interpolation used for missing values.

The results proved that such data-driven statistical approaches are
extremely useful in identifying unexpected and sometimes counter-intuitive
associations between the available spatial data nodes. It is worth
noticing that such discoveries verify and quantify important semantic
information that is related to special geographical features, such
as narrow sea passages and large obstacles to wind flows. This is
very important when designing corresponding models for short-term
forecasting of sea condition, especially average wave height and direction,
which is in fact what defines the associated weather risk of crossing
these passages in refugee influx patterns.

\appendices{}

\bibliographystyle{IEEEtran}
\bibliography{aegean-winds-grid}
\begin{IEEEbiography}{Harris Georgiou}
 received his B.Sc. degree in Informatics from University of Ioannina,
Greece, in 1997, and his M.Sc. degree in Digital Signal Processing
\& Computer Systems and Ph.D. degree in Machine Learning \& Medical
Imaging, from National \& Kapodistrian University of Athens, Greece,
in 2000 and 2009, respectively. Since 1998, he has been working as
an associate researcher, primarily with the Department of Informatics
\& Telecommunications at National \& Kapodistrian University of Athens
(NKUA/UoA), Greece. He has been actively involved in several national
and EU-funded research \& development projects, focusing on new and
emerging technologies in medical imaging and AI applications. He has
completed a 3-year post-doctorate research appointment with NKUA,
specializing in sparse learning models and fMRI/EEG signal for applications
in Biomedicine and Bioinformatics. Since 2016 he has been working
with the Informatics Dept. of Univ. of Piraeus on another post-doctorate
research appointment, focusing on emerging technologies and innovation
for next-gen air traffic management systems by trajectory prediction
via Machine Learning. He is also the active LEAR (team coordinator)
with the Hellenic Rescue Team of Attica (HRTA) on a `citizen observatories'
project for civil protection and flood events management via intelligent
crowd-sourcing and AI-enabled integration web platforms. Since 2004
he has been working in the private sector for many years as a consultant
in Software Engineering and Quality Assurance (EDP/IT), as well as
a professor in various ICT-related subjects. His main research interests
include Machine Learning, Pattern Recognition, Signal Processing,
Medical Imaging, Soft Computing and Game Theory. He has published
49 peer-reviewed journal \& conference papers plus 56 independent
\& open-access works, technical reports, magazine articles, software
toolboxes and open-access datasets, a two-volume book series on medical
imaging and diagnostic image analysis, contributed in three other
major books and one U.S. patent in related R\&D areas. He is a member
of the IEEE and the ACM organizations, general secretary in the A.C.
board of the Hellenic Informatics Union (HIU), team coordinator of
the Greek ICT4D task group \& the ``Sahana4Greece'' (virtual EOC
for the refugee crisis) initiative and he has given several technical
presentations in various countries.
\end{IEEEbiography}

\end{document}